\def\year{2021}\relax
\documentclass[letterpaper]{article} % DO NOT CHANGE THIS
\usepackage{aaai21}  % DO NOT CHANGE THIS
\usepackage{times}  % DO NOT CHANGE THIS
\usepackage{helvet} % DO NOT CHANGE THIS
\usepackage{courier}  % DO NOT CHANGE THIS
\usepackage[hyphens]{url}  % DO NOT CHANGE THIS
\usepackage{graphicx} % DO NOT CHANGE THIS
\usepackage{natbib}  % DO NOT CHANGE THIS AND DO NOT ADD ANY OPTIONS TO IT
\usepackage{caption} % DO NOT CHANGE THIS AND DO NOT ADD ANY OPTIONS TO IT
\usepackage{subcaption}
\usepackage{booktabs}

\usepackage{amsmath, amssymb}
\usepackage{multirow}
\usepackage{algorithm,algorithmic}
\usepackage{mathtools}
\usepackage{tikz}

\usepackage{amsthm}

\title{3DMolNet: A Generative Network for Molecular Structures}
\author{Vitali Nesterov, Mario Wieser, Volker Roth\\
}
 \affiliations{
 Department of Mathematics and Computer Science\\
 University of Basel, Switzerland\\
 \textit{vitali.nesterov@unibas.ch}\\
}

\begin{document}

\maketitle

\begin{abstract}
With the recent advances in machine learning for quantum chemistry, it is now possible to predict the chemical properties of compounds and to generate novel molecules. Existing generative models mostly use a string- or graph-based representation, but the precise three-dimensional coordinates of the atoms are usually not encoded. First attempts in this direction have been proposed, where autoregressive or GAN-based models generate atom coordinates. Those either lack a latent space in the autoregressive setting, such that a smooth exploration of the compound space is not possible, or cannot generalize to varying chemical compositions. We propose a new approach to efficiently generate molecular structures that are not restricted to a fixed size or composition. Our model is based on the variational autoencoder which learns a translation-, rotation-, and permutation-invariant low-dimensional representation of molecules. Our experiments yield a mean reconstruction error below 0.05 \AA, outperforming the current state-of-the-art methods by a factor of four, and which is even lower than the spatial quantization error of most chemical descriptors. The compositional and structural validity of newly generated molecules has been confirmed by quantum chemical methods in a set of experiments.
\end{abstract}

\begin{figure}[t]
\includegraphics[width=0.35\textwidth]{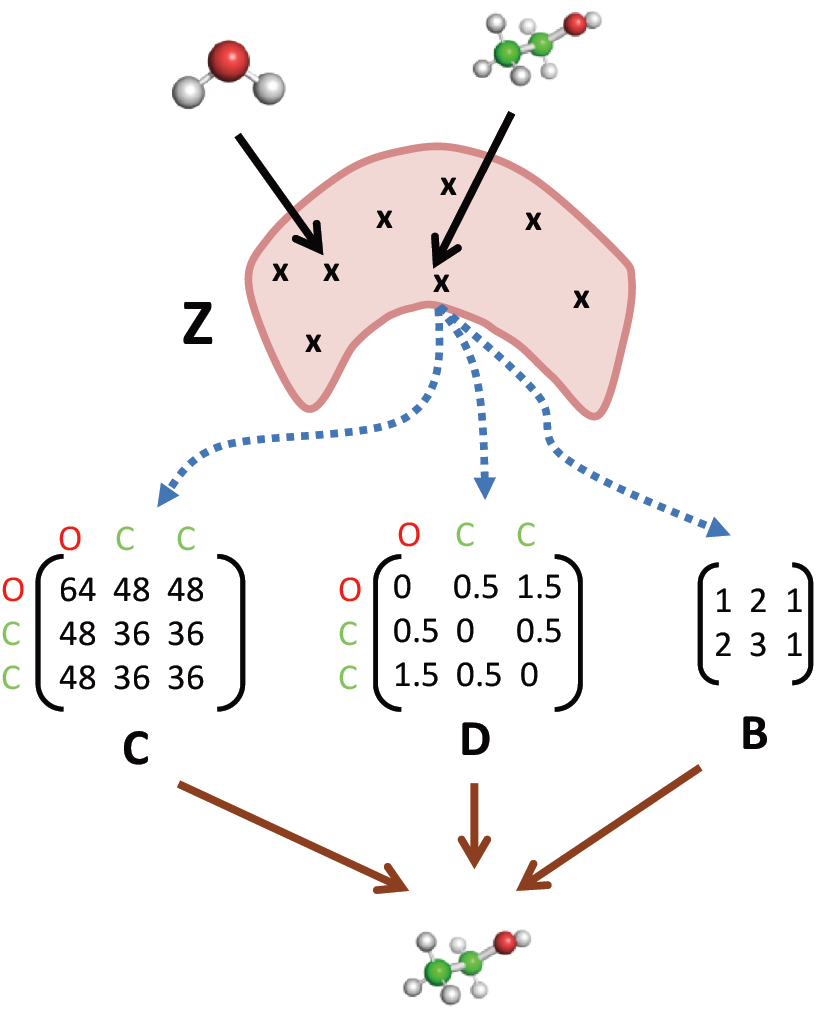}
\centering
\caption{A schematic illustration of the decoding process of the 3DMolNet. The model learns a low-dimensional representation $\mathbf{Z}$ of molecules (red plane). Each latent encoding can be decoded to a nuclear charge matrix $\mathbf{C}$, an Euclidean distance matrix $\mathbf{D}$, and a bond matrix $\mathbf{B}$ using neural networks (blue dotted arrows). From this, we recover the 3-d molecular structure of a molecule (brown arrows).}
\label{fig:gra_abstract}
\end{figure}

\section{Introduction}

An exploration of the chemical compound space (CCS) is essential in the design process of new materials. Considering that the amount of small organic molecules is in the magnitude of $10^{60}$ \cite{space}, a systematic coverage of molecules in a combinatorial manner is intractable. Furthermore, a precise approximation of quantum chemical properties by solving the Schr\"odinger's equation is computationally demanding even for a tiny amount of mid-sized molecules. In recent years, an entirely new research field emerged, aiming a statistical approach by training regression models for quantum chemical property prediction to overcome computational drawbacks \cite{schnet,faber2017prediction,christensen2019operators}. More recently, deep latent variable models have been introduced to solve the inverse problem and to generate novel molecules. Here, the major focus of the research community is on deep latent variable models that use either string- or graph-based molecular representations \cite{Gomez,kusner,jin18a,janz2017learning}. The precise spatial configuration of the molecules is usually not enclosed. This leads to various drawbacks, for instance, molecules which have the same chemical composition but different geometries, a so called isomers, cannot be distinguished. Furthermore, a targeted exploration of the CCS with respect to quantum chemical property is rather poor and with respect to geometries it is impossible.

To account for this challenges, \citeauthor{gebauer2019symmetry} proposed G-SchNet \cite{gebauer2019symmetry}, an autoregressive model for generation of atom types and corresponding 3-d points. The rotation- and translation-invariances are achieved by using Euclidean distances between atoms. The permutation-invariant features are extracted with continuous-filter convolutions \cite{schnet}. Due to the autoregressive nature of the model, it has no continuous latent space representation for smooth exploration of the CCS. Furthermore, the probability for completion of complex structures decays with the amount of sampling steps. \citeauthor{HoffmannNoe_EDMnets_2019} introduced EDMNet that generates Euclidean distance matrices (EDM) in a one-shot fashion, which can be further transformed into actual coordinates. For generation of the EDMs a GAN architecture \cite{goodfellow2014generative} is used, while the critic network is based on the SchNet architecture \cite{schnet}. This approach, however, is limited to a fixed chemical composition to circumvent the permutation problem. In addition, GANs are frequently difficult to train. A smooth exploration of the nearby areas in the latent space is not possible as well.

In order to overcome the aforementioned limitations, we propose the 3DMolNet, a generative network for 3-d molecular structures. A graphical sketch of our approach is illustrated in Figure \ref{fig:gra_abstract}. The molecule representation consists of two core components, a nuclear charge matrix and an EDM. Since some bond types have high variances in the bond lengths, an exact assignment of bond types often cannot be derived from the distances. For this reason, we additionally include an explicit bond matrix to our representation. To overcome the permutation problem, we use an InChI-based canonical ordering of the atoms. 

Due to the absence of canonical identifiers for most of the hydrogen atoms, we involve only heavy atoms in our representation. This is however not problematic, since the backbone of a molecule is the most important part. The hydrogens are explicitly defined if the core structure of a molecule is known. A quantum mechanical approximation of the hydrogen coordinates is computationally an easy optimization task. This is because, given a fixed core structure, a many-body problem has a significantly smaller degree of freedom. Additionally, the positions of the hydrogen atoms are entirely determined only if the stereospecificity is defined. 

Our model is based on the Variational autoencoder \cite{kingma2013auto,rezende14} architecture to learn a low-dimensional latent representation of the molecules. To recover a molecule, we decode each representation component separately and obtain the entire molecular structure in the post processing steps.

In summary, the main contributions of this work are:

\begin{enumerate}
\item We introduce a translation-, rotation-, and permutation-invariant molecule representation. To overcome the permutation problem a canonical ordering of the atoms is used.

\item We propose a VAE-based model to learn a continuous low-dimensional representation of the molecules and to generate 3-d molecular structures in a one-shot fashion.

\item We demonstrate on the QM9 dataset that the proposed model is able to reconstruct almost all chemical compositions with up to 9 heavy atoms. The reconstruction of coordinates for heavy atoms yields a RMSD below 0.05 \AA, outperforming the state-of-the-art method by almost a factor of four.

\end{enumerate}

\section{Related Work}

\paragraph{Variational Autoencoder.} 
In recent years, the Variational autoencoder (VAE) \cite{kingma2013auto,rezende14} became an important tool for various machine learning tasks such as semi-supervised learning \cite{NIPS2014_5352} and conditional image generation \cite{NIPS2015_5775}. An important line of research are VAEs for disentangled representation learning \cite{multi_level_vae,Klys,FaderNetworks,Creswell}. More recently, variational autoencoders have shown strong connections to the information bottleneck principle \cite{art:tishby:ib,AlemiFD016,infodropout}, where multiple model extensions have been proposed in the context of sparsity \cite{Wieczorek}, causality \cite{2018arXiv180702326P,mlhc}, archetypal analysis \cite{archetypes,keller2020learning} or invariant subspace learning \cite{Wieser2020Inverse}.

\paragraph{Graph-based Deep Generative Models.} Deep generative models are heavily used in the context of computational chemistry to generate novel molecular structures based on graph representations. With a text-based graph representation, a SMILES string, \cite{Gomez} introduced a chemical VAE to generate novel molecules. However, this method has limitations in generating syntactically correct SMILES strings. Several methods have been proposed to overcome this drawback \cite{kusner,sdvae,janz2017learning}. Another line of research deals with explicit graph-based molecular descriptors. \citeauthor{NIPS2018_8005} introduce a constrained graph VAE to generate valid molecular structures. Subsequently, improved approaches has been introduced by \cite{jin18a,Wengong,Yujia}.

\paragraph{Coordinate-based Deep Generative Models.} More recently, deep generative models for spatial structures has been proposed to overcome the limitations of graph-based descriptors. Such representations do not take into account precise three-dimensional configurations of the atoms. \citeauthor{Mansimov} estimated the 3-d configurations of molecules from molecular graphs. \citeauthor{HoffmannNoe_EDMnets_2019} learned the distribution of spatial structures by generating translation- and rotation-invariant Euclidean distance matrices. However, their approach is limited to molecules with the same chemical composition. Current state-of-the-art results are provided by the G-SchNet \cite{Gebauer}, which is based on an autoregressive model which, however, is lacking a latent representation of molecules and is inefficient and error-prone when sampling complex and large structures.

\section{Preliminaries}
\label{prelim}

\subsection{Variational Autoencoder}

The Variational autoencoder (VAE) \cite{kingma2013auto,rezende14} is a deep latent variable model that combines a generative process (decoder) with a variational inference (encoder) to learn a probabilistic model. The encoder models a posterior distribution $q(\boldsymbol{z} \mid \boldsymbol{x})$, where we assume a Gaussian distribution parametrized with $\boldsymbol{\mu_z}$ and $\boldsymbol{\sigma_z}$ and the decoder models the generative distribution $p(\boldsymbol{x} \mid \boldsymbol{z})$. In order to learn the probabilistic model, we optimize the following lower bound:

\begin{align}
L_{\mathrm{VAE}} &\geq \mathbb{E}_{\boldsymbol{z} \sim q(\boldsymbol{z} \mid \boldsymbol{x})}[\log p(\boldsymbol{x} \mid \boldsymbol{z})] \nonumber \\ &- D_{\mathrm{KL}}[q(\boldsymbol{z} \mid \boldsymbol{x}) \mid \mid p(\boldsymbol{z})].
\label{vae_bound} 
\end{align}

Due to the deterministic nature of neural networks, the random variable $Z$ is reparametrized as $z=\mu_z+\sigma_z*\epsilon$ where $\epsilon \sim \mathcal{N}(0,1)$.

\subsection{Quantum-mechanical Background}

A molecule as a quantum-mechanical system is fully described by the Schr\"odinger's (SE) equation. It is sufficient to consider a non-relativistic and time-independent formulation $H \Psi=E \Psi$. With an additional elimination of the external electromagnetic field, the Hamiltonian $H$ is determined by the external potential 

\begin{equation}
v(\mathbf{r})=\sum_{i} Z_{i} /\left|\mathbf{r}-\mathbf{R}_{i}\right|,
\label{eq:externalpot}
\end{equation}

where $\mathbf{Z} \in \mathbb{N}^{N}$ is a set of nuclear charges, $\mathbf{R} \in \mathbb{R}^{N \times 3}$ are the corresponding atomic positions, and $\mathbf{r} \in \mathbb{R}^{M \times 3}$ are stationary electronic states. The interatomic energy contributions result from the solution of the full electronic many-body problem and the nuclear Coulomb repulsion energy: 

\begin{equation}
Z_{i} Z_{j} /\left|\mathbf{R}_{i}-\mathbf{R}_{j}\right|.
\label{eq:repulsionenergy}
\end{equation}

Neglecting the former, a physical system depends only on a set of nuclear charge numbers and the corresponding atom coordinates in a three-dimensional space.

\section{Methods}

The goal of our learning framework is to capture the distribution of the molecular structures, which are expressed in terms of chemical compositions and geometries. To overcome the permutation problem, we use a canonical ordering of the atoms. In our approach we exclude the hydrogen atoms from the representation, since those can be easily added and the coordinates can be cheaply optimized with quantum-mechanical calculations. Our model involves a VAE-based architecture for smooth exploration of the chemical space. To enforce the quality of the generated EDMs, additional geometry loss terms are added. The molecular core structure is reconstructed from the estimated representation components. The entire molecule is recovered in the post processing steps.

\subsection{Representation}

A physical many-body system depends only on a set of nuclear charge numbers $Z_i \in \mathbb{N}$ and corresponding $\mathbf{R}_i \in \mathbb{R}^{3}$ coordinates. Our molecule representation involves this information with a nuclear charge matrix $\mathbf{C} \in \mathbb{N}^{N \times N}$ and an Euclidean distance matrix $\mathbf{D} \in \mathbb{R}^{N \times N}$. Additionally, a bond matrix $\mathbf{B} \in \mathbb{N}^{M \times 3}$ is included to encode explicit bond types. We define the nuclear charge matrix as follows:

\begin{equation}
C_{i j} = Z_i Z^{\top}_j.
\label{eq:chargematrix}
\end{equation}

The distances matrix is defined with

\begin{equation}
D_{i j}=\left\|\boldsymbol{R}_{i} - \boldsymbol{R}_{j} \right\|_{2}^{2}.
\label{eq:chargematrix}
\end{equation}

The bond matrix $\mathbf{B}$ includes indices of connected atoms in the first and the second columns. The third column contains bond type values. Although the bond matrix could be extracted from the distances, the variance of the bond lengths relatively high for some bond types, such that an assignment from the distances is often ambiguous. To handle molecules of different sizes, the charge, distance, and bond matrices are padded with zero rows and or columns to a fixed size. The final molecule representation is a set of the three components:

\begin{equation}
\mathbf{X} = \left\{ \mathbf{C}, \mathbf{D}, \mathbf{B} \right\}.
\label{eq:rep}
\end{equation}

\subsection{Canonical Identifier}

A major obstacle for a learning algorithm is an arbitrary atom ordering. To overcome this problem we generate a unique atom ordering for molecules. There exist different canonicalization algorithms. We get the canonical identifiers (CIs) with an InChI-based algorithm implemented in the CDK package \cite{cdk}. The generation of CIs involves a structure normalization, an InChI-based canonical labelling, and a tree traversal \cite{o2012towards}. With this method, we get CIs for each heavy atom and only for explicit hydrogen atoms. Those are indicated isotops, dihydrogen and hydrogen ions, and hydrogen atoms attached to tetrahedral stereocentres with defined stereochemistry. Due to the absence of canonical labels for most of the hydrogen atoms, our representation involves only heavy atoms.

\begin{figure*}[!h]
\includegraphics[width=\textwidth]{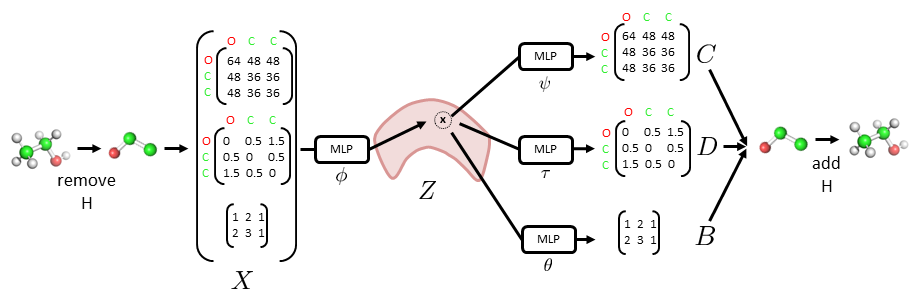}
\centering
\caption{An illustration of the 3DMolNet framework. In the preprocessing step, we apply a canonical ordering and remove all hydrogen atoms. We then get the representation $\mathbf{X}$ as a concatenation of the nuclear charge $\mathbf{C}$, the euclidean distance $\mathbf{D}$, and the bond $\mathbf{B}$ matrices. Subsequently, we map $\mathbf{X}$ into a low-dimensional latent representation $\mathbf{Z}$ using an encoder neural network which is parametrized with $\phi$. Here, (x) denotes the mean and the surrounding circle the variance of our estimate. To decode a molecule, we use three separate neural networks parametrized with $\psi, \tau$, and $\theta$ which decode representation components $\mathbf{C}$, $\mathbf{D}$, and $\mathbf{B}$ separately. In the post processing step, we recover the chemical and structural composition of the heavy atoms with a multidimensional scaling algorithm and add hydrogen atoms with Open Babel \cite{OBoyle2011a} and fine-tune the coordinates with MOPAC \cite{mopac} while fixing the positions of heavy atoms.}
\label{fig:model}
\end{figure*}

\subsection{Model}

As previously stated, our aim is to generate novel molecular compositions and corresponding 3-d structures (see Figure \ref{fig:gra_abstract}). Therefore, we want to learn both, a low-dimensional and a rotation-, translation-, and permutation-invariant representation of molecules (see Figure \ref{fig:model}). To do so, we reformulate the standard VAE framework (Equation \ref{vae_bound}) by defining each representation component $\mathbf{C}$, $\mathbf{D}$, and $\mathbf{B}$ in a form of independent random variables. This leads to an extended parametric formulation given with:

\begin{align}
 L_{\mathrm{VAE}} &\geq \mathbb{E}_{\boldsymbol{z} \sim q(\boldsymbol{z} \mid \boldsymbol{c},\boldsymbol{d},\boldsymbol{b})}[p(\boldsymbol{c},\boldsymbol{d},\boldsymbol{b} \mid \boldsymbol{z})]\\\nonumber &- \beta D_{\mathrm{KL}}[q(\boldsymbol{z} \mid \boldsymbol{c},\boldsymbol{d},\boldsymbol{b}) \mid \mid p(\boldsymbol{z})].
\label{3dmolnet_bound} 
\end{align}

\paragraph{Encoder.} The encoder is defined as the KL-divergence between the posterior $q_\phi(\boldsymbol{z} \mid \boldsymbol{c},\boldsymbol{d},\boldsymbol{b})$ and the prior $p(\boldsymbol{z})$ and is denoted as:

\begin{equation}
    D_{\mathrm{KL}}[q_\phi(\boldsymbol{z} \mid \boldsymbol{c},\boldsymbol{d},\boldsymbol{b}) \mid \mid p(\boldsymbol{z})]\nonumber,
\end{equation}

where $\phi$ are the neural network parameters. We define the posterior as a Gaussian distribution and assume a Gaussian prior $p(\boldsymbol{z})=\mathcal{N}(0,\boldsymbol{I})$.

\paragraph{Decoder.} The decoder is defined as the the negative log-likelihood of $p(\boldsymbol{c},\boldsymbol{d},\boldsymbol{b} \mid \boldsymbol{z})$. As we assume conditional independence between $\boldsymbol{c}$, $\boldsymbol{d}$, and $\boldsymbol{b}$ we can express the joint distribution as follows:

\begin{align}
    \mathbb{E}_{\boldsymbol{z} \sim q(\boldsymbol{z} \mid \nonumber \boldsymbol{c},\boldsymbol{d},\boldsymbol{b})}[p(\boldsymbol{c},\boldsymbol{d},\boldsymbol{b} \mid \boldsymbol{z})] &= \mathbb{E}_{\boldsymbol{z}_\phi \sim q(\boldsymbol{z} \mid \boldsymbol{c},\boldsymbol{d},\boldsymbol{b})}[p_\theta(\boldsymbol{c} \mid \boldsymbol{z})\\\nonumber
    &\cdot p_\tau(\boldsymbol{d} \mid \boldsymbol{z}) \\\nonumber
        &\cdot p_\psi(\boldsymbol{b} \mid \boldsymbol{z})],
\end{align}

where $\phi$, $\tau$ and $\theta$ denote neural network parameters of each respective decoder. We define each log-likelihood term to be the mean-absolute-error between a particular representation component and the reconstructed counterpart.

\paragraph{Geometry Loss.} 

To further improve the quality of the EDM reconstruction, we additionally penalize the negative eigenvalues and the rank of the Gram matrix. To do so, we first define the geometric centering matrix

\begin{equation}
\boldsymbol{J} = \boldsymbol{I} - \frac{1}{N} \mathbf{1} \mathbf{1}^{\top},
\label{eq:centering}
\end{equation}

where $\boldsymbol{I}$ is the identity matrix and $\mathbf{1}$ is the all-ones vector. The Gram matrix is defined with

\begin{equation}
\boldsymbol{G} = -\frac{1}{2} \boldsymbol{J} \boldsymbol{D} \boldsymbol{J}.
\label{eq:gram}
\end{equation}

The matrix $\boldsymbol{D}$ is an EDM, if the Gram matrix $\boldsymbol{G}$ is positive semi-definite. To enforce this property we penalize negative eigenvalues of the Gram matrix $\boldsymbol{G}$. Therefor, we first get eigenvalues $\boldsymbol{\lambda}$ with the eigenvalue decomposition of $\boldsymbol{G}$. We then apply the ReLU function to the negative of the eigenvalues $\bar{\boldsymbol{\lambda}} = \operatorname{ReLU}(-\boldsymbol{\lambda})$. The corresponding loss term is defined as follows:

\begin{equation}
L_{\mathrm{EV}} = \bar{\boldsymbol{\lambda}}^{\top} \bar{\boldsymbol{\lambda}}.
\label{eq:loss:ev}
\end{equation}

Since the atom coordinates exists in a maximally 3-dimensional embedding, we additionally penalize larger than $k = 3$ rank of the Gram matrix $\boldsymbol{G}$. For this, we sort the eigenvalues in descending order $\boldsymbol{\lambda} = [\lambda_{1} \leq \lambda_{2} \leq \dots \lambda_{N}]$. The rank loss is obtained with

\begin{equation}
L_{\mathrm{R}} = \sum_{i = k + 1}^{N} \lambda_{i}^{2}.
\label{eq:negeval}
\end{equation}

\paragraph{Overall Training Objective.} 

The total loss function is given with:

\begin{equation}
    L_{\mathrm{total}} = L_{\mathrm{VAE}} + L_{\mathrm{EV}} + L_{\mathrm{R}}.
\end{equation}

\subsection{Recovering Molecular Structures}

After having introduced our model, we now describe the mechanism to get the entire molecule from the nuclear charge, distance, and bond matrices. To obtain the atom and coordinate pairs we first symmetrize the generated nuclear charge matrix $\boldsymbol{C}$ and the distance matrix $\boldsymbol{D}$. The diagonal of the distance matrix $\boldsymbol{D}$ is set to zero. The reconstructed floating point values of the bond matrix $\boldsymbol{B}$ are rounded to integer values. Subsequently, we recover a set of the nuclear charge numbers and the corresponding coordinates $\left\{Z_{i}, \mathbf{R}_{i}\right\}$. To do so, we use the classical multidimensional scaling algorithm. We then use Open Babel to read-in the molecular structure by setting types and coordinates of the atoms and assign bonds and corresponding bond types. Lastly, we reconstruct a complete molecule by adding hydrogen atoms with Open Babel. To get initial positions of hydrogen atoms we use an efficient force-filed method with Open Babel \cite{OBoyle2011a} and fine-tune by using a semi-empirical \textit{ab initio} approximation with MOPAC \cite{mopac}. 

Since the ordering of the added hydrogen atoms differ from the ordering in the initial molecule, the best matching identity assignment has to be found in order to meaningfully calculate the deviation of the structures. Given the coordinates of the added and the target hydrogen atoms, we use the Hungarian algorithm \cite{kuhn1955hungarian} to minimize the assignment costs.

\section{Experiments and Discussion}

\subsection{Dataset}
For our experiments, we use the QM9 dataset \cite{rama2014}, which includes 133,885 small organic molecules and consist of up to nine heavy atoms (C, O, N, and F). Each molecule has geometric, energetic, electronic, and thermodynamic properties obtained from the Density Functional Theory (DFT) calculations. The chemical space of QM9 is based on the GDB-17 dataset \cite{gdb17}, enumerating more then 166 billion molecules of up to 17 heavy atoms. GDB-17 is systematically generated based on the molecular connectivity graphs and approaches a complete and unbiased enumeration of the chemical compound space of small and stable organic molecules. However, the molecules are generated based on a set of fixed rules, those does not cover all valid molecular structures which could be uncovered with deep generative models.

\subsection{Experimental Setup} 
We train the 3DMolNet on a set of 50K randomly selected molecules from the QM9 dataset. For validation 5K molecules are randomly selected. The remaining molecules are used for evaluation as a test set. We use Open Babel for validity check of the valences. We also remove molecules from the QM9 dataset which are inconsistent in terms of the provided bonds in the SMILES representation of the molecule and bonds assigned by Open Babel. To compare molecular geometries, we apply Procrustes analysis and calculate the root-mean-square deviation (RMSD) of pair-wise atomic coordinates between generated and ground truth structures. To compare with G-SchNet, we used code and a trained model provided under a published source.

\subsection{Architecture} 
3DMolNet is based on a classical variational autoencoder architecture. It consists of a single encoder and three decoder networks for generation of the nuclear charge matrix $\boldsymbol{C}$, the distance matrix $\boldsymbol{D}$, and the bond matrix $\boldsymbol{B}$. The encoder and each decoder have a standard fully connected architecture. The latent space is set to 64 dimensions. The compression parameter $\beta$ is reduced after each epoch.

\begin{table}
\caption{A summary of quantitative results for reconstruction of heavy atoms, as well as the entire molecular structures including hydrogen atoms. The error is reported in the RMSD metrics as a median of pair-wise atomic distances between the ground-truth and geometrically aligned reconstructs.}
\centering
 \begin{tabular}{@{}lrrrr@{}}
 \toprule
Model & \multicolumn{4}{c}{Reconstruction Accuracy}\\
\cmidrule(r{1em}){2-5} 
&\multirow{1}{1.3cm}{\centering heavy} &\multirow{1}{1.3cm}{\centering all} \\
\midrule
Mansimov & -\hspace{3 mm} & 0.37\hspace{3 mm} \\
G-SchNet & 0.18\hspace{3 mm} & 0.21\hspace{3 mm} \\
\midrule
3DMolNet & \textbf{0.05}\hspace{3 mm} & \textbf{0.16}\hspace{3 mm} \\
\bottomrule
\end{tabular}
\label{tab:modelcomp}
\end{table}

\subsection{Reconstruction Accuracy} 
In order to judge about the quality of the 3DMolNet, we evaluate the reconstruction accuracy of the nuclear charge numbers, the atom coordinates, and the bond matrix. Our experiments show that the chemical compositions are exactly reconstructed in more than 99 \% of the cases. The bond matrix is exactly reconstructed in 98 \% of the cases. To evaluate the reconstructed geometries, we only accept molecules with exactly reconstructed atoms, bonds, and bond types. The reconstruction of heavy atom coordinates yields a RMSD of 0.048 \AA. By using a force-field for optimization of the coordinates of the hydrogen atoms, a RMSD of 0.21 \AA\ is achieved. Applying a further optimization step by using a semi-empirical \textit{ab initio} method, the RMSD drops to 0.16 \AA\ (see Table \ref{tab:modelcomp}).

To compare with G-SchNet we first investigate the reconstruction accuracy of the chemical compositions. According to our experiments, the chemical compositions are exactly reconstructed in only 19 \% of the cases. Even though this result is obtained with teacher forcing as a one-step-ahead prediction, it still significantly lacks behind our accuracy rates. According to published numbers, G-SchNet yields a RMSD of 0.18 \AA\ for the reconstruction of heavy atom coordinates. This is roughly four times as high compared to our results. By including the hydrogen atoms, G-SchNet achieves a RMSD of 0.23 \AA\, however using a fast force-field solution for optimizing hydrogen coordinates, we achieve comparable results. By using a more precise approximation, our method clearly outperforms G-SchNet. 

A comparison with EDMNet in terms of the reconstruction accuracy of the coordinates cannot be directly done, due to the nature of the chosen GAN-based architecture. Beyond that, a fair comparison is not possible anyway, since the EDMNet is trained only on a sub set of molecules of the same chemical composition.

\subsection{Interpolation between Molecules} 

An important and useful property of our model is a continuous and smooth latent space, which allows a gradual exploration of the nearby chemical compositions and structures. In a row of experiments, we evaluate and discuss the outcomes of an interpolation between molecular domains. Our start and target molecules are chosen from the test set. To estimate the quality of the sampled structures, we additionally relax the entire geometry of the molecules in each interpolation step.

\begin{figure}[!h]
    \centering
        \begin{subfigure}{0.19\textwidth}
        		\centering
        		\includegraphics[width=\textwidth]{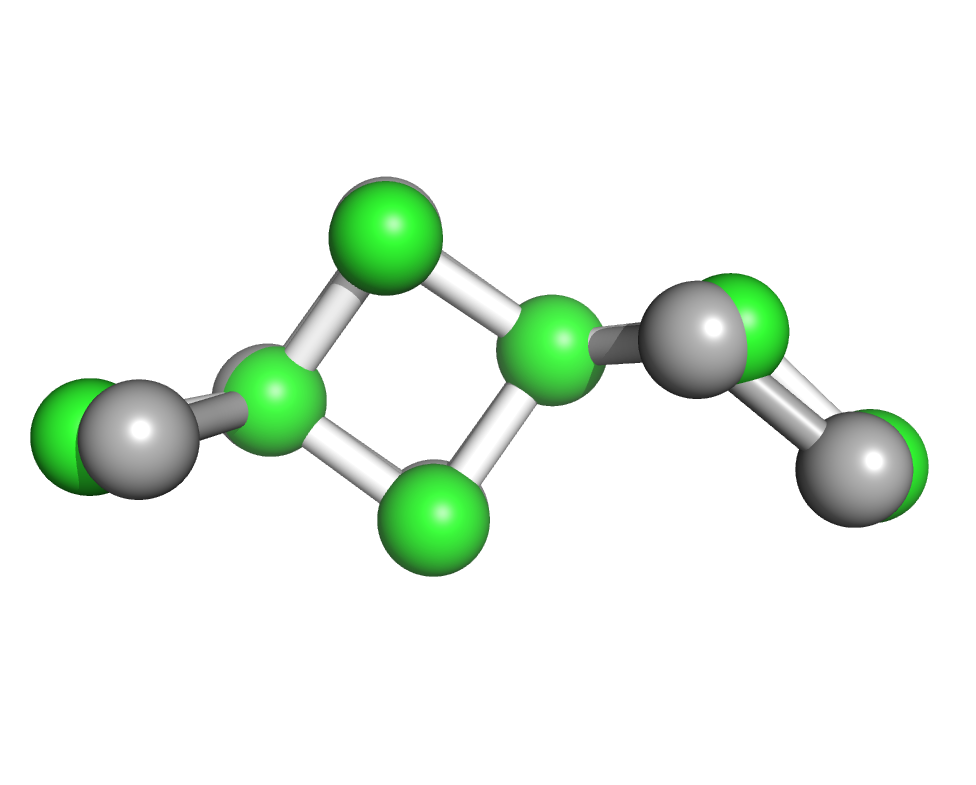}
        \end{subfigure}   \hspace{4mm}%
        \begin{subfigure}{0.19\textwidth}
                \centering
        		\includegraphics[width=\textwidth]{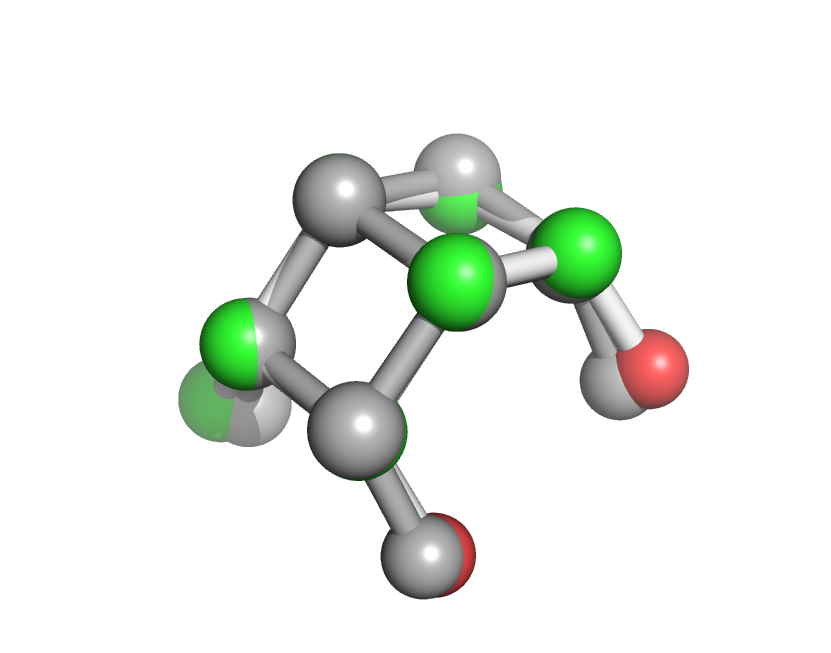}
        \end{subfigure} 
        \begin{subfigure}{0.19\textwidth}
                \centering
        		\includegraphics[width=\textwidth]{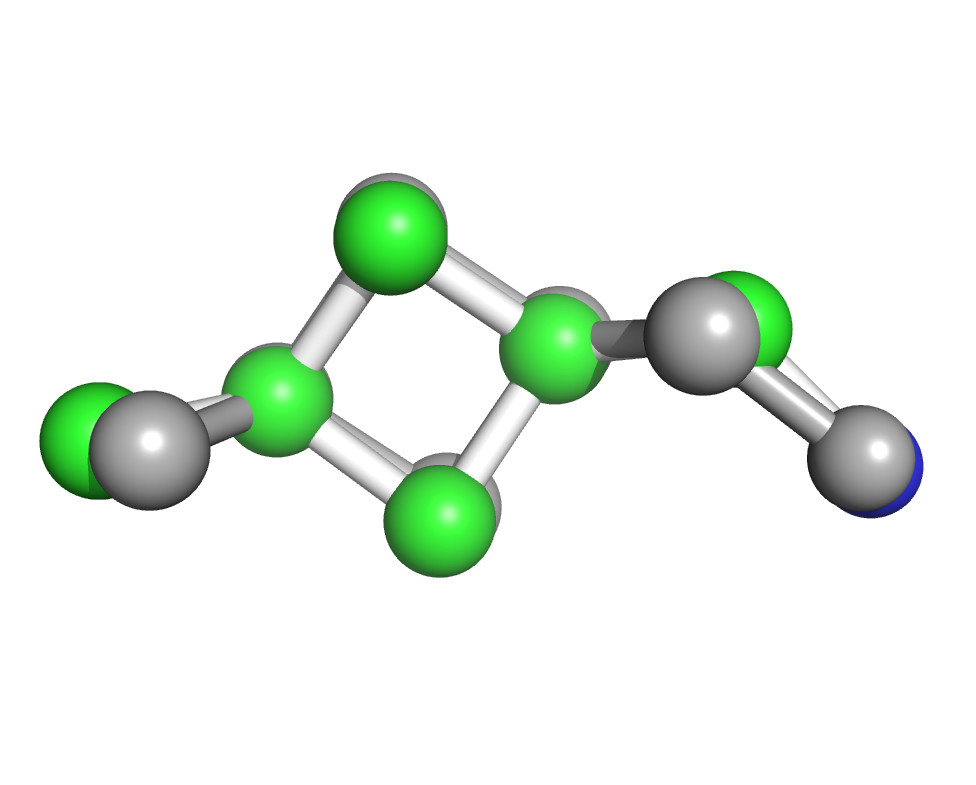}
        \end{subfigure}
        \begin{subfigure}{0.19\textwidth}
                \centering
        		\includegraphics[width=\textwidth]{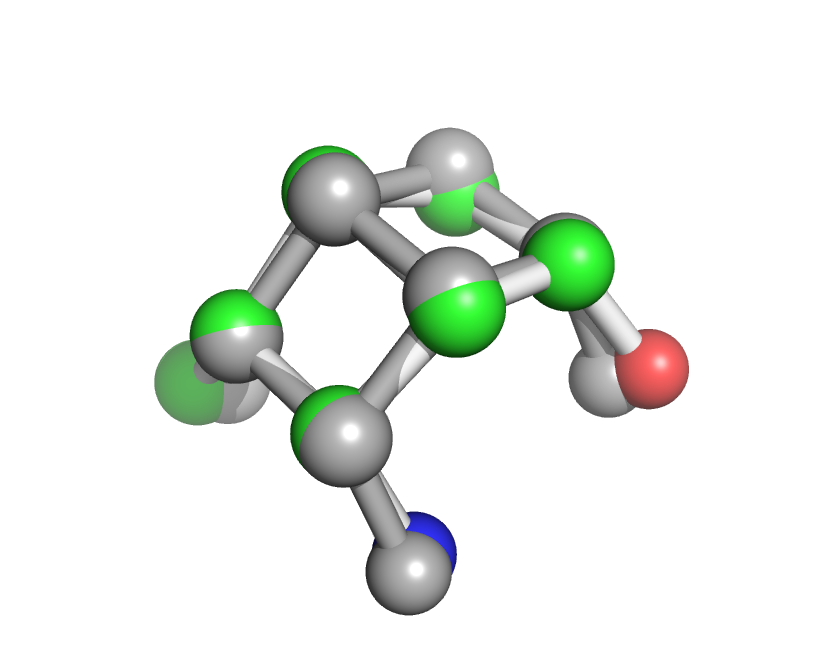}
        \end{subfigure} 
        \begin{subfigure}{0.19\textwidth}
                \centering
        		\includegraphics[width=\textwidth]{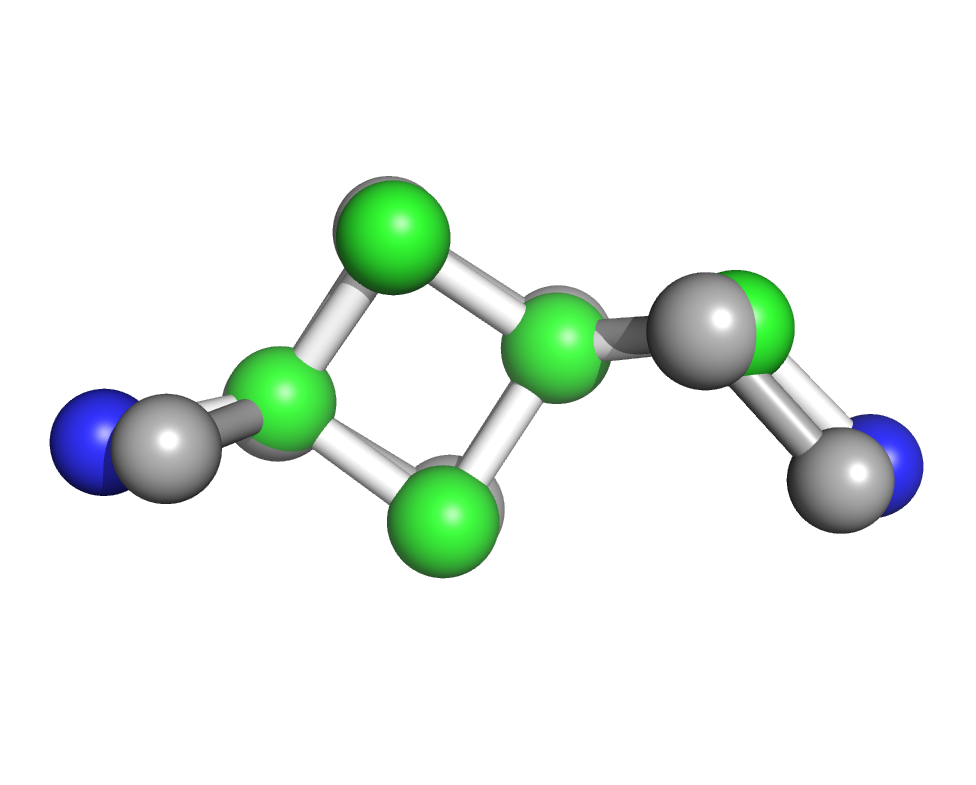}
        \end{subfigure}
        \begin{subfigure}{0.19\textwidth}
                \centering
        		\includegraphics[width=\textwidth]{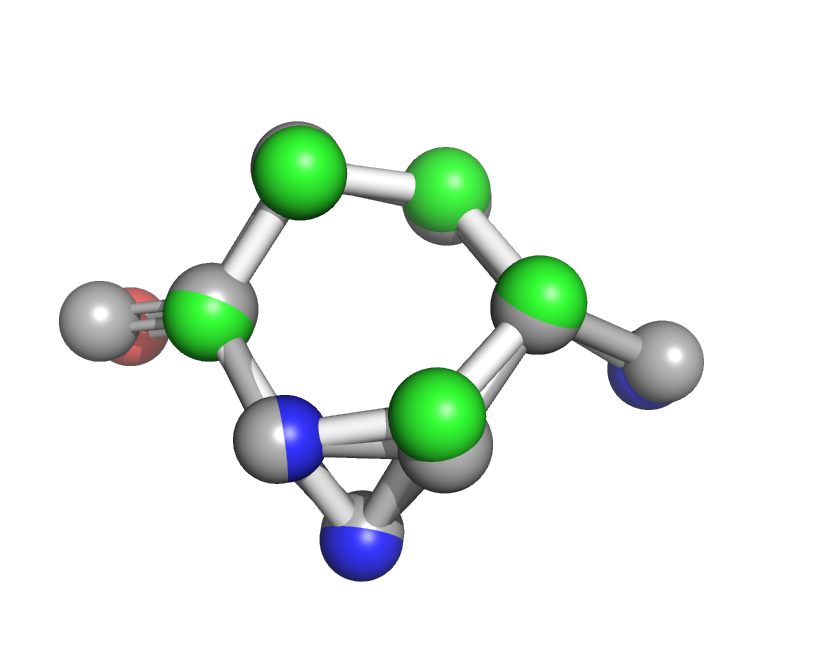}
        \end{subfigure} 
        \begin{subfigure}{0.19\textwidth}
                \centering
        		\includegraphics[width=\textwidth]{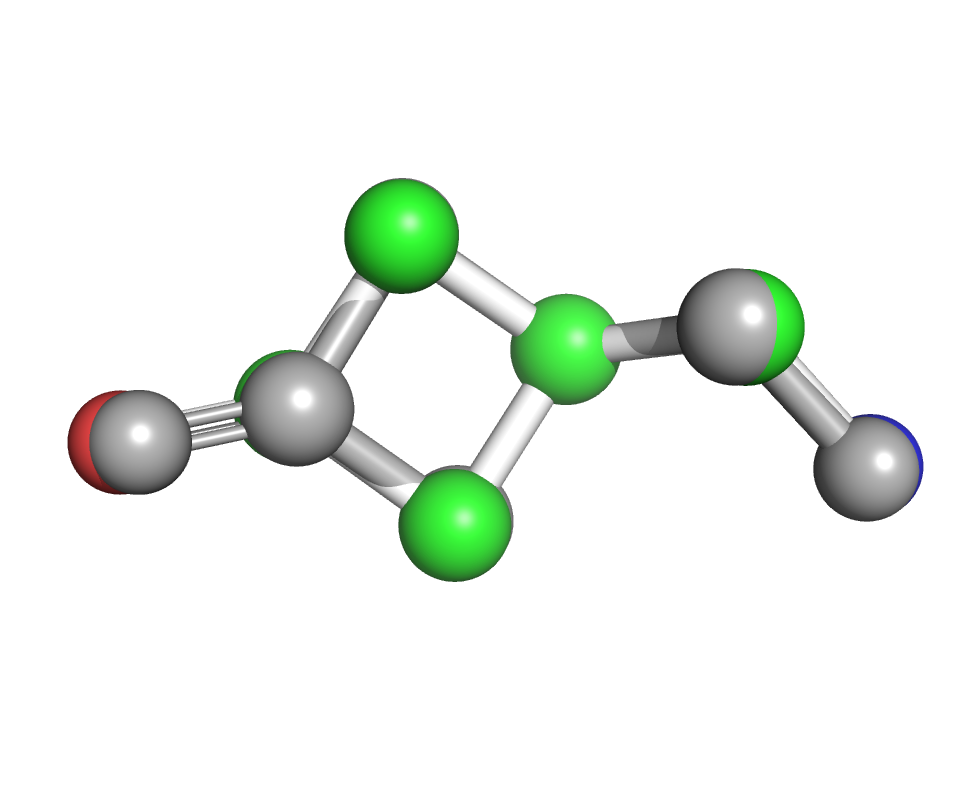}
        \end{subfigure} 
        \begin{subfigure}{0.19\textwidth}
                \centering
        		\includegraphics[width=\textwidth]{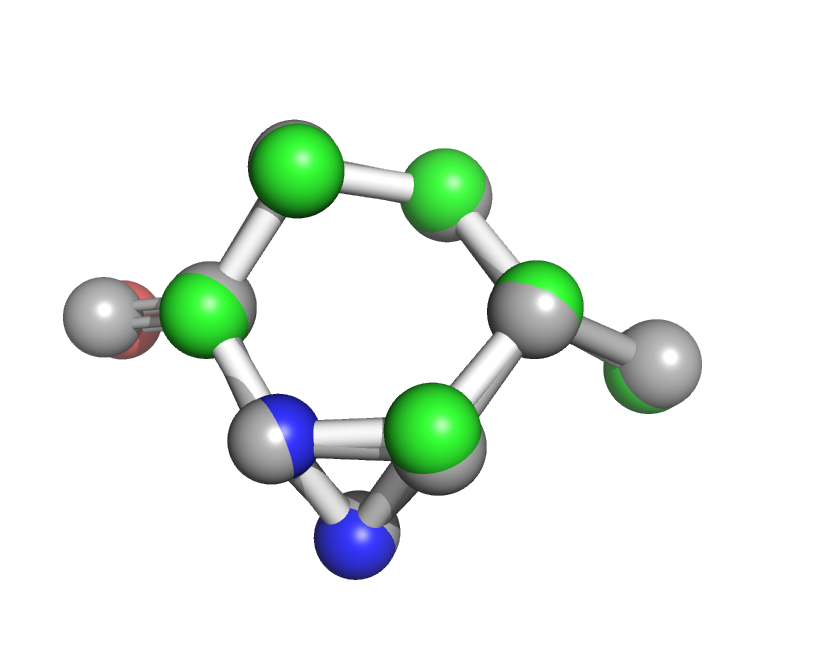}
        \end{subfigure} 
        \begin{subfigure}{0.19\textwidth}
                \centering
        		\includegraphics[width=\textwidth]{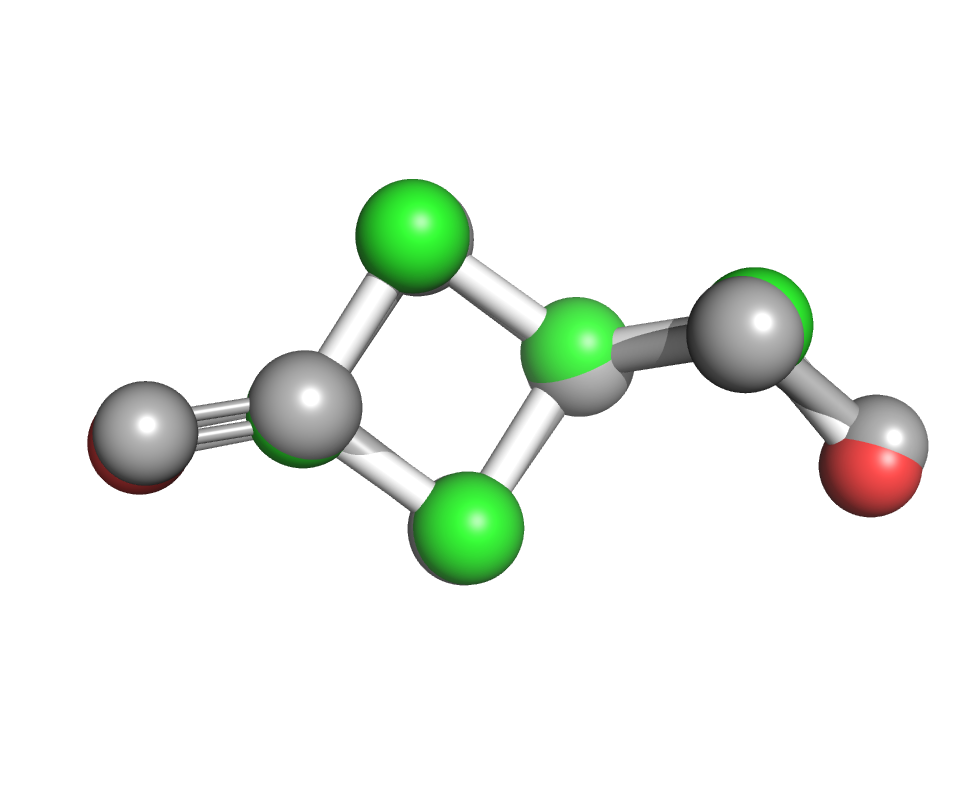}
        		\caption{}
        		\label{fig:interpola}
        \end{subfigure} 
         \begin{subfigure}{0.19\textwidth}
                \centering
        		\includegraphics[width=\textwidth]{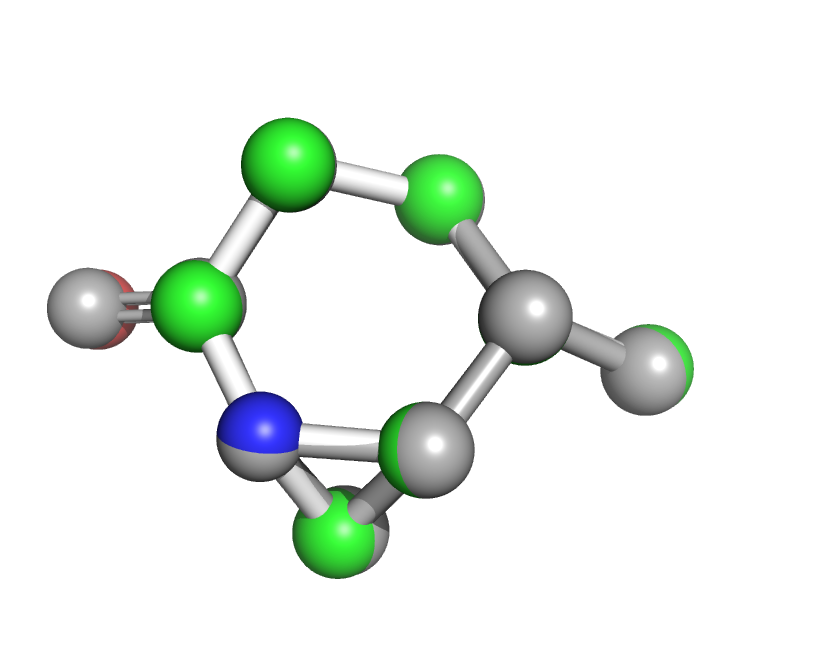}
        		\caption{}
        		\label{fig:interpolb}
        \end{subfigure} 
        \caption{An illustration of the interpolation steps between related molecules. The colored atoms and bonds represent generated molecules, while gray structures represent relaxed counterparts. The hydrogen atoms are removed. (a) depicts an interpolation between different chemical compositions within a fixed structural domain, while (b) shows an interpolation between different geometries as well.}
\label{fig:interpol}
\end{figure}

In general, we could observe smooth transitions between molecular structures and chemical compositions, meaning that structure- and composition-related molecules are mapped close to each other in the latent space, i.e the latent space is structured with respect to compositional features. Figure \ref{fig:interpola} depicts an example of a smooth transition between structurally related molecules, yet consisting of different atom types. Here, the first and the last molecule are recovered from the mean of data points picked out of the test set. By moving towards the target molecule, we plot each intermediate molecule when ever the decoded chemical composition changes. Additional gray plots depict corresponding relaxed geometries of the molecules.

In the second example (see Figure \ref{fig:interpolb}), the chemical composition, as well as the molecular geometry changes, as the interpolation takes place between less related molecules. During the interpolation process we discovered molecular geometries which closely correspond with the relaxed counterparts. However, we also found molecules which deviate from the relaxed geometries. The most common differences occurs within the reconstructs of the hydrogen coordinates. Further deviations are caused by core sub-structures of the molecule, naturally leading to significantly different outcomes for the hydrogen coordinates. This phenomenon is more closely discussed in the next paragraph. 

A direct comparison of our results with the G-SchNet and EDMNet is at this point not possible, since those models do not allow a continuous exploration of the latent space. Hence, an interpolation between molecular domains cannot be performed. In case of the EDMNet, even a transition between different chemical compositions and molecule sizes is not possible.

\begin{figure*}[!h]
    \centering
        \begin{subfigure}{0.20\textwidth}
        		\centering
        		\includegraphics[width=\textwidth]{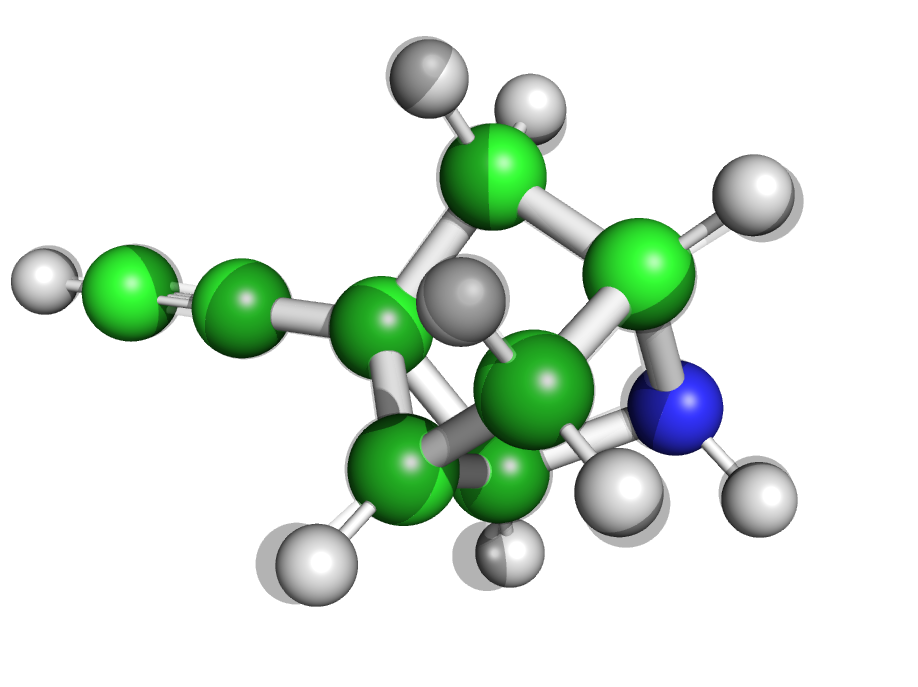}
        		\label{anamodel}
        		\caption{0.05}
        \end{subfigure}   \hspace{4mm}%
        \begin{subfigure}{0.20\textwidth}
        		\centering
        		\includegraphics[width=\textwidth]{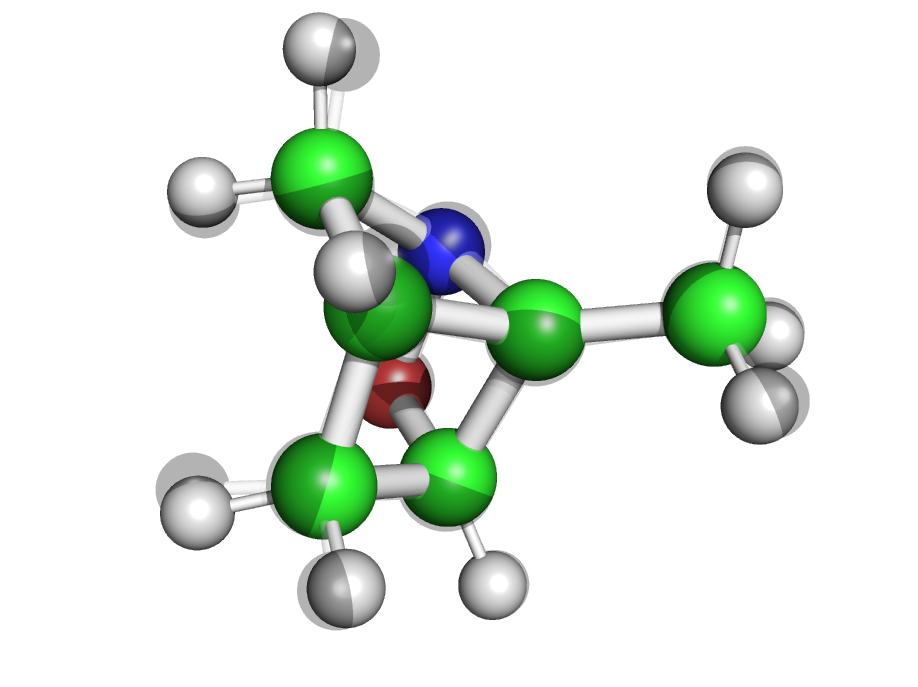}
        		\label{anamodel}
        		\caption{0.07}
        \end{subfigure}   \hspace{4mm}%
        \begin{subfigure}{0.20\textwidth}
                \centering
        		\includegraphics[width=\textwidth]{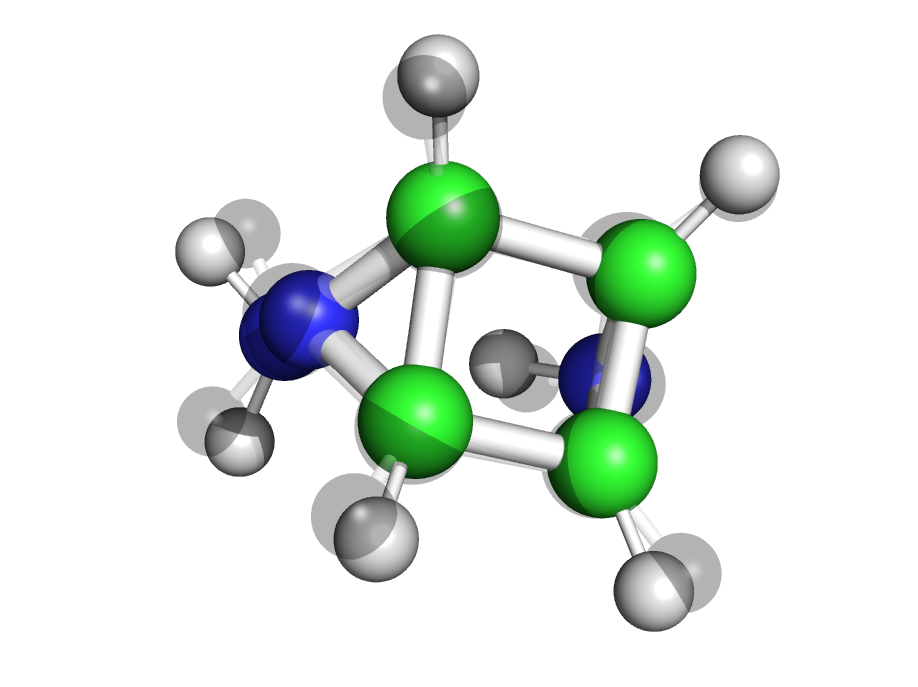}
        		\label{ourmodel}
        		\caption{0.12}
        \end{subfigure}
        \begin{subfigure}{0.20\textwidth}
                \centering
        		\includegraphics[width=\textwidth]{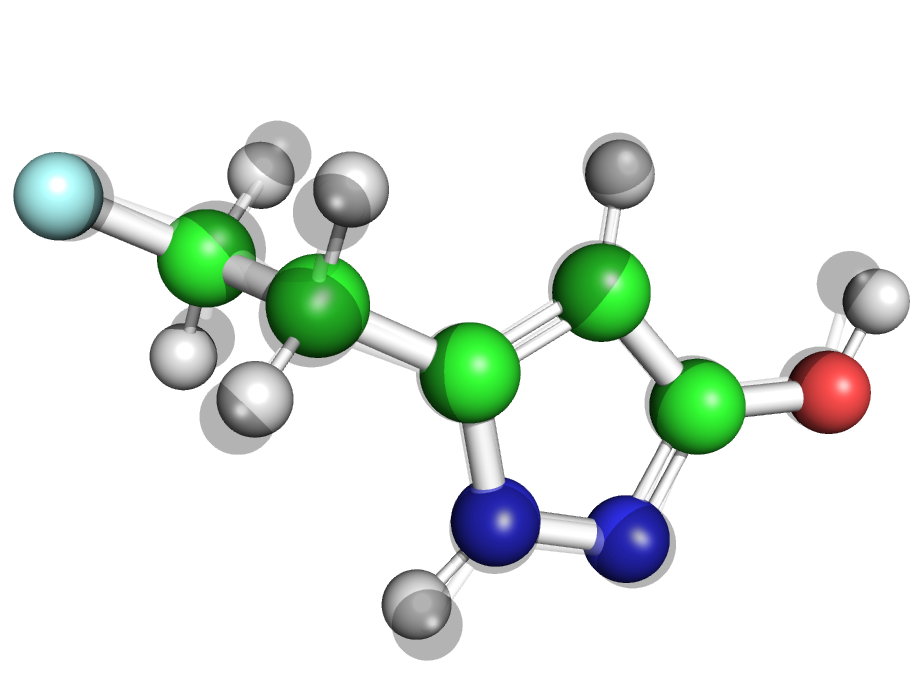}
        		\label{ourmodel}
        		\caption{0.17}
        \end{subfigure}   \hspace{4mm}%
        \begin{subfigure}{0.20\textwidth}
        		\centering
        		\includegraphics[width=\textwidth]{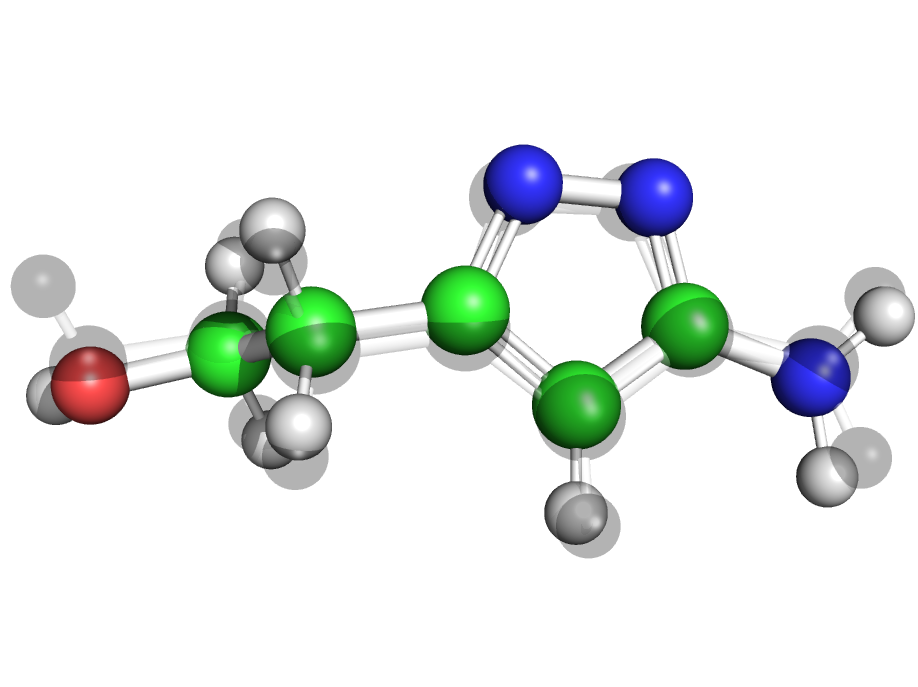}
        		\label{anamodel}
        		\caption{0.24}
        \end{subfigure}   \hspace{4mm}%
        \begin{subfigure}{0.20\textwidth}
                \centering
        		\includegraphics[width=\textwidth]{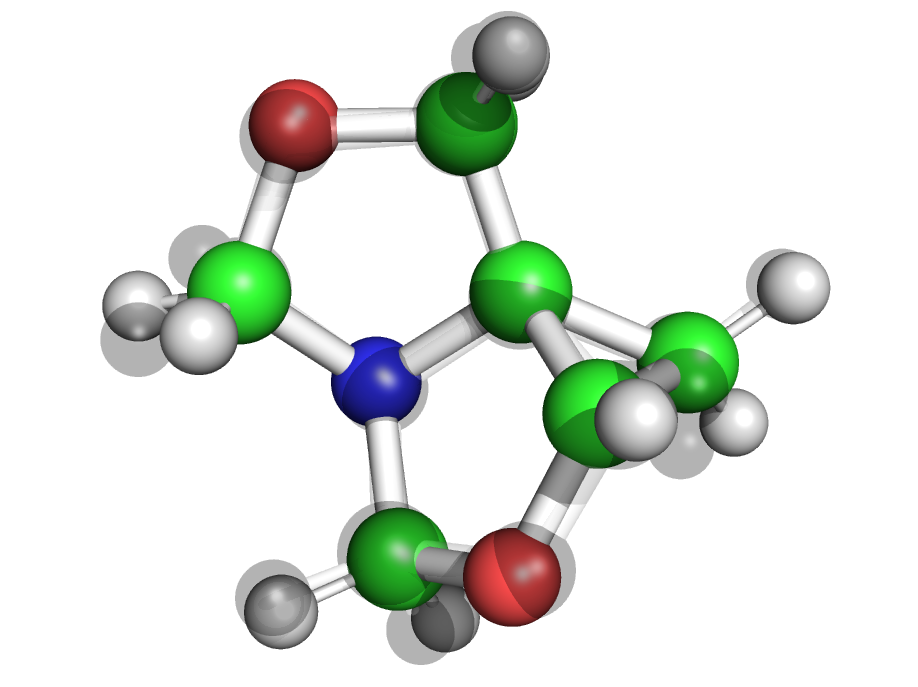}
        		\label{ourmodel}
        		\caption{0.39}
        \end{subfigure}
        \begin{subfigure}{0.20\textwidth}
        		\centering
        		\includegraphics[width=\textwidth]{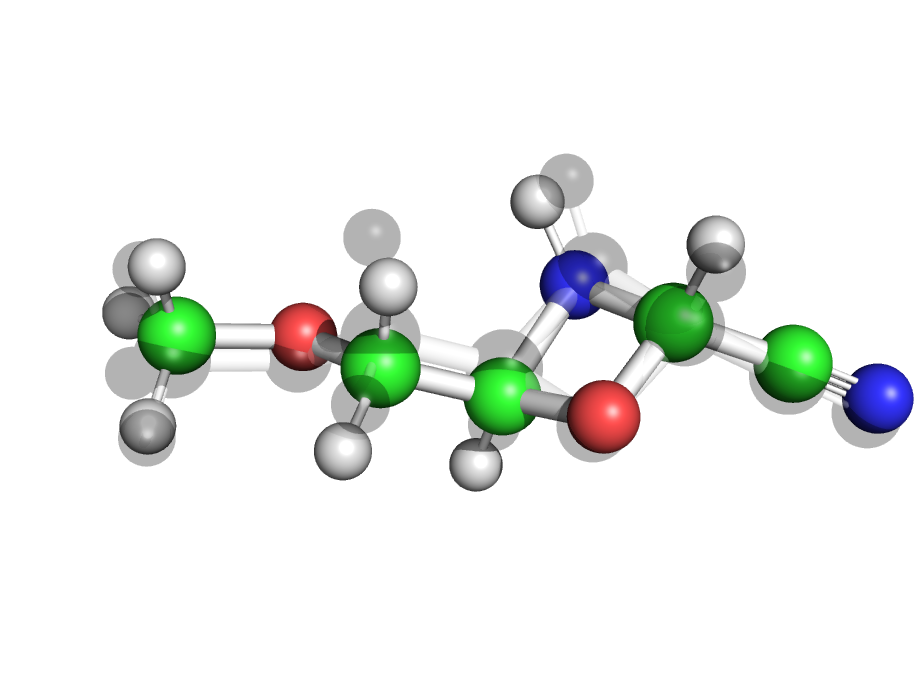}
        		\label{anamodel}
        		\caption{0.42}
        \end{subfigure}   \hspace{4mm}%
        \begin{subfigure}{0.20\textwidth}
                \centering
        		\includegraphics[width=\textwidth]{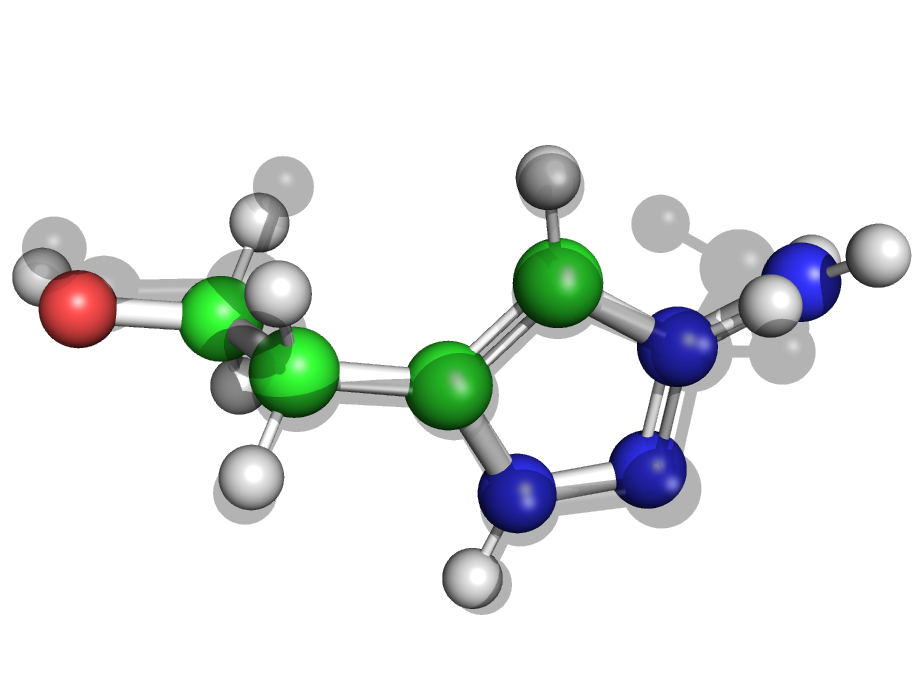}
        		\label{ourmodel}
        		\caption{0.55}
        \end{subfigure}
        \caption{An illustration of novel molecules discovered with 3DMolNet. The colored atoms and bonds represents generated structures and the translucent plot is the relaxed counterpart. The molecules are labeled with the RMSD for the relaxation.}
\label{fig:novel}
\end{figure*}

\subsection{Molecule Discovery} 

For discovery of novel molecules within the QM9 dataset, we randomly sample from the latent space around the mean regions of the molecules from the training set. To identify a novel chemical composition, first, we generate a set of canonical SMILES strings for the entire QM9 dataset. We then compare the canonical SMILES string of the generate molecule with the QM9 references. A molecule is accepted and is considered as novel if no identical canonical SMILES string is found within the QM9 dataset. Molecules with invalid bond and bond types are rejected. We were able to identify more than 20K novel molecules with new chemical compositions (see Figure \ref{fig:novel}). To investigate the quality of the generated molecular structures we relaxed the structures with MOPAC and computed the pair-wise atom distances yielding in a RMSD of 0.32 \AA.

In Figure \ref{fig:novel}, we illustrate a selection of the discovered molecules with increasing RMSD values between the generated structures and corresponding equilibrium states. One sees that the heavy atoms deviate the least from the ground-truth. The hydrogen atoms contribute most to the deviation. However, from the chemical perspective the exact reconstruction of the hydrogen atoms is least important, since most of them are not restricted to a fixed configuration. In terms of qualitative exploration of the chemical space, heavy atoms contribute most to the structural features of the molecules.

To put our results into relation, for G-SchNet a median of around 0.3 \AA\ is reported, being a comparable result achieved in our experiments. However, a direct comparison between G-SchNet and our model is not appropriate, since our molecules are sampled from a continuous latent space which can produce distorted and unstable molecular structures as well. In fact, it turned out to be problematic to fairly reason about the structural closeness to the equilibrium conformations of the relaxed counterparts. Due to a high-dimensional latent space it is computationally demanding to find mean coordinates within a latent space domain of discovered molecules. This is, however, important in order to decode the least distorted version of the molecule. Furthermore, some novel molecules could be within a domain of isomers, i.e being in a family of different molecular structures with the same chemical composition, which are mapped closely in the latent space. Hence, sampling in the nearby regions can generate considerably different relaxations and worsen the results. Additionally, generated molecules require a further criterion to filter out potential chemically unstable structures or even to penalize those reconstructions during the training. \\

\section{Conclusion}

We have developed the 3DMolNet to efficiently generate novel 3-d molecular structures of a variable size and chemical composition. In comparison, few recent methods are either autoregressive or GAN-based. The former lacks a latent space entirely, while the latter is restricted due to GAN-specific drawbacks, such that both models do not allow a smooth exploration of the chemical space. Additional restrictions of the GAN-based model allow only generation of fixed chemical compositions. 

To address these limitations, we make three distinct contributions:
\begin{enumerate}

\item We introduced a translation-, rotation-, and permutation-invariant molecule representation involving a canonical ordering of the atom and coordinate pairs.

\item We proposed an extended version of the Variational autoencoder, which allow to efficiently generate 3-d molecular structures and explore molecular domains within a continuous low-dimensional representation of the molecules.

\item We achieved a high reconstruction precision of the atom coordinates, which is below 0.05 \AA\ and is in the range of a typical spatial quantization error of common chemical descriptors. Furthermore, our model almost perfectly reconstructs exact chemical compositions and bond types on a QM9 test set.
\end{enumerate}

\subsection{Future Work} 
In future work we want to improve the quality of the geometries by including molecular stability constraints to generate energetically more reasonable configurations. Furthermore, we want to extend the model with a decoder for hydrogen atoms. Lastly, to improve the process of the latent space exploration, we would like to investigate alternative forms of latent representations.

\subsection{Acknowledgments}

We would like to thank Anders S. Christensen, Felix A. Faber, Puck van Gerwen, O. Anatole von Lilienfeld, and Jimmy C. Kromann for insightful comments and helpful discussions. Vitali Nesterov is partially supported by the NCCR MARVEL funded by the Swiss National Science Foundation. Mario Wieser is partially supported by the NCCR MARVEL and grant 51MRP0158328 (SystemsX.ch HIV-X) funded by the Swiss National Science Foundation.

\small{
    \bibliography{bib}
    \bibliographystyle{aaai}
}

\end{document}